\documentclass[lettersize,journal]{IEEEtran}
\usepackage{amsmath,amsfonts}
\usepackage{algorithmic}
\usepackage{algorithm}
\usepackage{array}
\usepackage[caption=false,font=normalsize,labelfont=sf,textfont=sf]{subfig}
\usepackage{textcomp}
\usepackage{stfloats}
\usepackage{url}
\usepackage{verbatim}
\usepackage{graphicx}
\usepackage{cite}
\usepackage{color}
\usepackage{soul}
\usepackage{ulem}
\usepackage{upgreek} 
\begin{document}

\title{Nuclear Magnetic Resonance Measurements in High Flat-top Pulsed Magnetic Field up to 40 T at WHMFC}

\author{Wenqi Wei, Qinying Liu, Le Yuan, Jian Zhang, Shiyu Liu, Rui Zhou, Yongkang Luo, and Xiaotao Han
\thanks{This work was supported in part by the National Key Research and Development Plan Project of China under Grant 2021YFA1600301 and 2022YFA1602602, in part by the National Natural Science Foundation of China under Grant U21A20458 and 51821005, and in part by the Interdisciplinary Program of Wuhan National High Magnetic Field Center under Grant WHMFC202126. (Corresponding author: Xiaotao Han.)}
\thanks{Wenqi Wei, Le Yuan, Jian Zhang, Shiyu Liu, Yongkang Luo, and Xiaotao Han are with the Wuhan National High Magnetic Field Center, Huazhong University of Science and Technology, Wuhan 430074, China (e-mail: vinkey7wei@hust.edu.cn; mpzslyk@gmail.com; xthan@hust.edu.cn).}
\thanks{Qinying Liu is with the Electric Power Research Institute, China Southern Power Grid Company Limited, Guangzhou 510623, China (e-mail: liuqy1@csg.cn).}
\thanks{Rui Zhou is with the Institute of Physics, Chinese Academy of Sciences, and Beijing National Laboratory for Condensed Matter Physics, Beijing 100190, China (e-mail: rzhou@iphy.ac.cn).}}


\maketitle

\begin{abstract}
Nuclear magnetic resonance (NMR) technique benefits from the high magnetic field not only due to the field-enhanced measurement sensitivity and resolution, but also because it is a powerful tool to investigate field-induced physics in modern material science. In this study, we successfully performed NMR measurements in the high flat-top pulsed magnetic field (FTPMF) up to 40 T. A two-stage corrected FTPMF with fluctuation of less than 10 mT and duration of longer than 9 ms was established. Besides, a Giga-Hz NMR spectrometer and a sample probe suitable for the pulsed-field condition were developed. Both free-induction-decay and spin-echo sequences were exploited for the measurements. The derived $^{93}$Nb NMR results show that the stability and homogeneity of the FTPMF reach an order of 10$^2$ ppm / 10 ms and 10$^2$ ppm / 10 mm$^3$ respectively, which is approaching a degree of maturity for some researches on condensed matter physics.
\end{abstract}


\section{Introduction}
\IEEEPARstart{E}{ver} since discovered in the 1940s, nuclear magnetic resonance (NMR) has manifested itself as one of the most significant tools to derive detailed information on the atomic scale about material properties. Nowadays, the trend of performing NMR experiments in the high magnetic field is actively driven not only by higher sensitivity and resolution, but also by the desirable exploration of ﬁeld-induced physics in modern material science\cite{ref1,ref2,ref3,ref4,ref5,ref6,ref7,ref8,ref9,ref10,ref11,ref12,ref13,ref14,ref15,ref16,ref17,ref18}. For instance, researches on half-integer quadrupolar nuclei such as $^{27}$Al and $^{17}$O notably benefit from the high magnetic field, because the quadrupolar interaction induced shift and broadening of the central transition are inversely proportional to the square of the magnetic field strength. The reduction in second-order quadrupolar broadening brings resolution enhancement and additional information, which provides new opportunities for measurements of nuclei not probed before\cite{ref1,ref2,ref3,ref4,ref5,ref8,ref10}. On the other side, since the high magnetic field often causes electronic and structural transitions of materials, the interest in field-induced physics like high-$T_{\rm c}$ cuprate superconductors has been increasing in the community of modern condensed matter physics. The NMR spectroscopy can provide more fundamental insights in the properties of materials, hence people have been making efforts to carry out NMR experiments in the higher magnetic field\cite{ref6,ref7,ref9,ref11,ref12,ref13,ref14,ref15,ref16,ref17,ref18}.

At present, the high magnetic field of greater than 30 T can be generated by continuous-operation (also called “steady-state” or “DC”) or pulsed magnets. The DC magnets with operation time of more than seconds include a variety of water-cooled resistive magnets (up to 41.5 T\cite{ref19}), superconducting magnets (up to 32.35 T\cite{ref20}), and hybrid magnets (up to 45.5 T\cite{ref21}). The continuous magnetic field is applicable to perform NMR experiments, because there is enough time to correct the field both in stability and homogeneity, as well as to execute various sequences for measurements and achieve signal averaging. However, large construction and operation costs and limited field strength of superconducting conductors hinder the development of NMR experiments to higher DC magnetic field.

Currently, only pulsed resistive magnets are practically available to produce the high magnetic field exceeding 45 T and even up to 100 T for timescales typically in the range of 1-100 ms\cite{ref22,ref23}. Although some NMR experiments were successfully optimized and achieved in the short and time-varying pulsed field\cite{ref6,ref7,ref8,ref9,ref10,ref11,ref12,ref13,ref14,ref15,ref16}, it is still desirable to perform NMR measurements in a relatively stable magnetic field\cite{ref17,ref18,ref24}. Firstly, a large real-time bandwidth up to tens of MHz or higher is required to excite and detect NMR signals due to the low repeatability and time-varying of the pulse field, which brings great difficulties to hardware design and signal search\cite{ref25}. Besides, the instability of the pulsed magnetic field causes the violent fluctuation of the resonance frequency and leads to the distortion of the NMR spectra. Despite the application of complex signal post-correction algorithms, the signal quality is still limited\cite{ref13,ref15}. Furthermore, some highly accurate parameter measurements such as the relaxation time $T_1$ and $T_2$ demand that the time dependence of the magnetic field should be reduced to zero during long enough sequence durations\cite{ref24}. Moreover, the eddy current induced by the time-varying field may create additional magnetic field gradients that degrade the NMR spectral resolution, and can heat the metal sample further causing the measurement error. Hence, it is still challenging to carry out NMR measurements in the pulsed magnetic field.

Fortunately, lots of efforts have been being made to create a quasi-stable stage (usually called “flat-top”) near the peak of the pulsed magnetic field\cite{ref26,ref27,ref28,ref29,ref30,ref31,ref32}, opening a new avenue for the pulsed magnetic field NMR measurements\cite{ref18}. In the early stage, Weickert $et$ $al.$\cite{ref26} tried to use a long-pulse magnet powered by high-power capacitor banks to create an approximate 44 T flat-top for NMR experiments. However, the unregulated long pulsed magnetic field is inefficient in improving the flat-top stability and has a long repetition time of more than 8 hours. Recently, Ihara and Kohama $et$ $al.$\cite{ref17,ref24} systematically performed NMR measurements in a dynamically controlled flat-top field, showing the great advantages of the flat-top pulsed magnetic field (FTPMF). Nevertheless, their measurements are limited in the low field of 13 T, and a higher FTPMF is desired in the community.

In this work, we present a practically available FTPMF for NMR measurements up to 40 T at Wuhan National High Magnetic Field Center (WHMFC). In Sec. II, we briefly introduce the basic concepts of NMR measurement. Furthermore, experimental setups are presented in Sec. III including a two-stage corrected flat-top pulsed magnetic field, a modular GHz NMR spectrometer, and a sample probe. In Sec. IV, NMR measurements of $^{93}$Nb carried out in the DC field and FTPMF are described and discussed. Finally, the paper is concluded in Sec. V.

\section{Basic Concepts of NMR Measurement}
To understand the basic concepts of NMR measurement and the influences of temporal instability and spatial inhomogeneity of the magnetic field on the measurement, a brief description is presented here and shown in Fig. 1. The detailed principles of NMR measurement can be found in many classic textbooks\cite{levitt2013spin,slichter2013principles}. The interaction between an external magnetic field  $\boldsymbol{B}_0$ and nuclear spin $I$ ($I>0$) is described by the Zeeman effect, in which the spectral line splits into multiple discrete levels. For the simplest case of $I=\frac{1}{2}$, the Zeeman energy levels are shown in Fig. 1(a) and are given by
\begin{equation}
E_{\mathrm{Zee}}=\pm \frac{1}{2}h\gamma B_0
\end{equation}
where $\gamma$ is gyromagnetic ratio (in MHz/T), and $h$ is Planck's constant. The difference of energy levels can be written as
\begin{equation}
\varDelta E_{\mathrm{Zee}}=hf_{\mathrm{L}}=h\gamma B_0
\end{equation}
where $f_{\mathrm{L}}$ is called the Larmor frequency and $f_{\mathrm{L}}=\gamma B_0$. If a radio-frequency (RF) field $\boldsymbol{B}_1\left( t \right) =B_1\cos \left( 2\pi f_{\mathrm{L}}t \right)$ perpendicular to the $\boldsymbol{B}_0$ excites the nucleus,  transitions between these energy levels can occur and NMR phenomenon can be observed.

Commonly, a ${\pi /2}$  RF pulse is used to irradiate the NMR free induction decay (FID) signals which are presented in Fig. 1(b) and have the form of 
\begin{equation}
s_{\mathrm{FID}}\left( t \right) \sim \cos \left( 2\pi f_{\mathrm{L}}t \right) \cdot e^{-\frac{t}{T_{2}^{*}}}
\end{equation}
where $T_{2}^{*}$ is the decay time. It is noteworthy that there is an approximately linear dependence of  NMR sensitivity on the strength of the ${B}_0$ \cite{ref4}, thus the high magnetic field is beneficial to the NMR measurement.

In the corresponding Fourier transform spectrum as shown in Fig. 1(c), the standard NMR spectrum is of the following form
\begin{equation}
S\left( f \right) \sim \frac{1/T_{2}^{*}}{\left( 1/T_{2}^{*} \right) ^2+\left( 2\pi f-2\pi f_L \right) ^2}
\end{equation}
which is called the absorption Lorentzian. It is obvious that the peak of the spectrum is at the Larmor frequency $f_{\mathrm{L}}$.

\begin{figure}[t]
\centering
\includegraphics[width=3.3in]{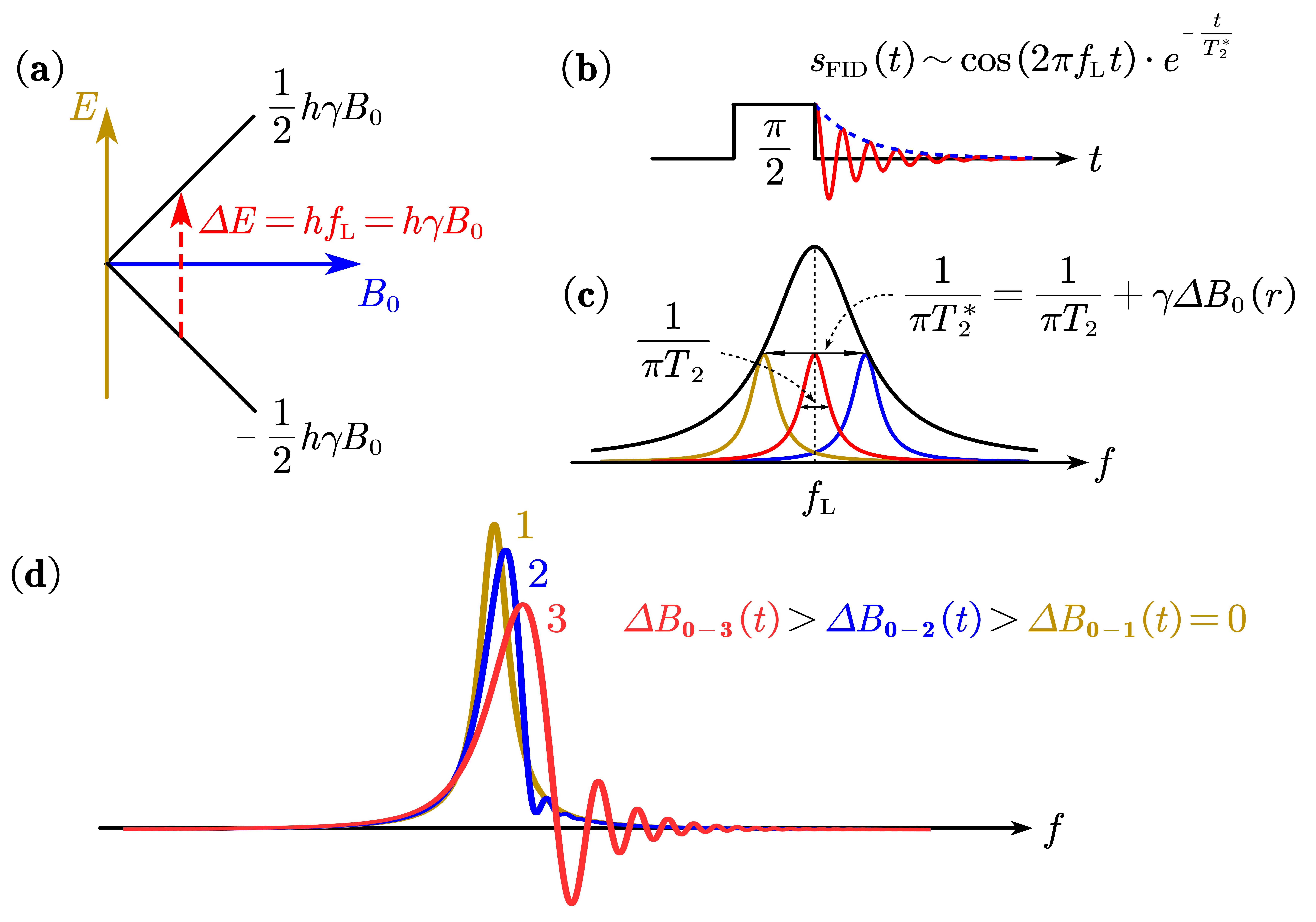}
\caption{Basic principles of NMR measurement. (a) The Zeeman energy levels of a nucleus with spin 1/2 in the presence of an external magnetic field $B_0$. NMR transition occurs at the Larmor frequency $f_{\mathrm{L}}=\gamma B_0$. (b) FID signals irradiated by a ${\pi /2}$ RF pulse. (c) A standard absorption Lorentzian NMR spectrum. the FWHM of $1/\left( \pi T_{2}^{*} \right)$ includes the intrinsic homogeneous broadening quantified by the transverse relaxation time $T_2$ and the inhomogeneous broadening due to the spatial inhomogeneity $\varDelta B_0\left( r \right)$ of the $B_0$. (d) The distortion of the NMR spectrum due to the temporal instability of the $B_0$. The $\varDelta B_0\left( t \right)$ reflects the degree of instability.}
\end{figure}

For the conventional NMR measurement in the pulsed magnetic field, when the external magnetic field $B_0$ changes with the time during the period of the FID, the $f_{\mathrm{L}}$ also changes proportionally and this leads to the distortion of the NMR spectrum \cite{ref8,ref15}. A simulation illustration is presented in Fig. 1(d) to compare the NMR responses under different magnetic field stability, and the distortion is positively correlated with the temporal instability of the $B_0$. Hence, high stability of the $B_0$ with time is one of the key prerequisites for the NMR measurement and the FTPMF is desired.

Another essential parameter in the NMR spectrum is the full width at half maxima (FWHM) parameterized by $1/\left( \pi T_{2}^{*} \right)$, which represents the spectral resolution. In fact, the FWHM includes the intrinsic homogeneous broadening quantified by the transverse relaxation time $T_2$ and the inhomogeneous broadening due to the spatial inhomogeneity $\varDelta B_0\left( r \right)$ of the $B_0$ across sample volume \cite{chavhan2009principles}
\begin{equation}
\frac{1}{\pi T_{2}^{*}}=\frac{1}{\pi T_2}+\gamma \varDelta B_0\left( r \right).
\end{equation}

Consequently, $B_0$ with less spatial inhomogeneity is desired for higher spectral resolution. Usually, samples positioned precisely with small volume are adopted to reduce the spectrum broadening caused by the magnetic field inhomogeneity in the pulsed magnet.

\section{Experimental Setups}
\subsection{Flat-top Pulsed Magnetic Field}
For decades, the FTPMF technique has been pursued to improve the stability and duration of the flat-top stage while maintaining the high field strength\cite{ref26,ref27,ref28,ref29,ref30,ref31,ref32}. The FTPMF is mainly powered by three types of power sources: the flywheel generator, the lead-acid battery bank, and the capacitor bank. The flywheel generator is the most ideal power source for generating the FTPMF with high strength and long duration up to 60 T/ 100 ms\cite{ref30}. However, it has a high cost of construction, operation and maintenance, which limits its application. The lead-acid battery bank can be regulated to generate the FTPMF with high stability and long duration up to 23 T ± 2 mT/ 100 ms\cite{ref31}, but its small supply voltage leads to the low strength of the FTPMF. The economical and reliable high-voltage capacitor bank is the most commonly used main power source for generating the high pulsed magnetic field, but it is difficult to regulate the discharge current for controlling the magnetic field. Despite all this, some solutions have been proposed, such as sequentially discharging multiple capacitors\cite{ref27,ref28,ref32}, or using a small compensation coil built into the bore of the main magnet to finely regulate the magnetic field\cite{ref29}. The former method can achieve a long flat-top duration of more than 10 ms but has high instability beyond 10$^3$ ppm, while the latter stands the opposite. Hence, we combined the advantages of the two schemes, and the flat-top realization is divided into two processes as shown in Fig. 2.

\begin{figure}[h]
\centering
\includegraphics[width=3.4in]{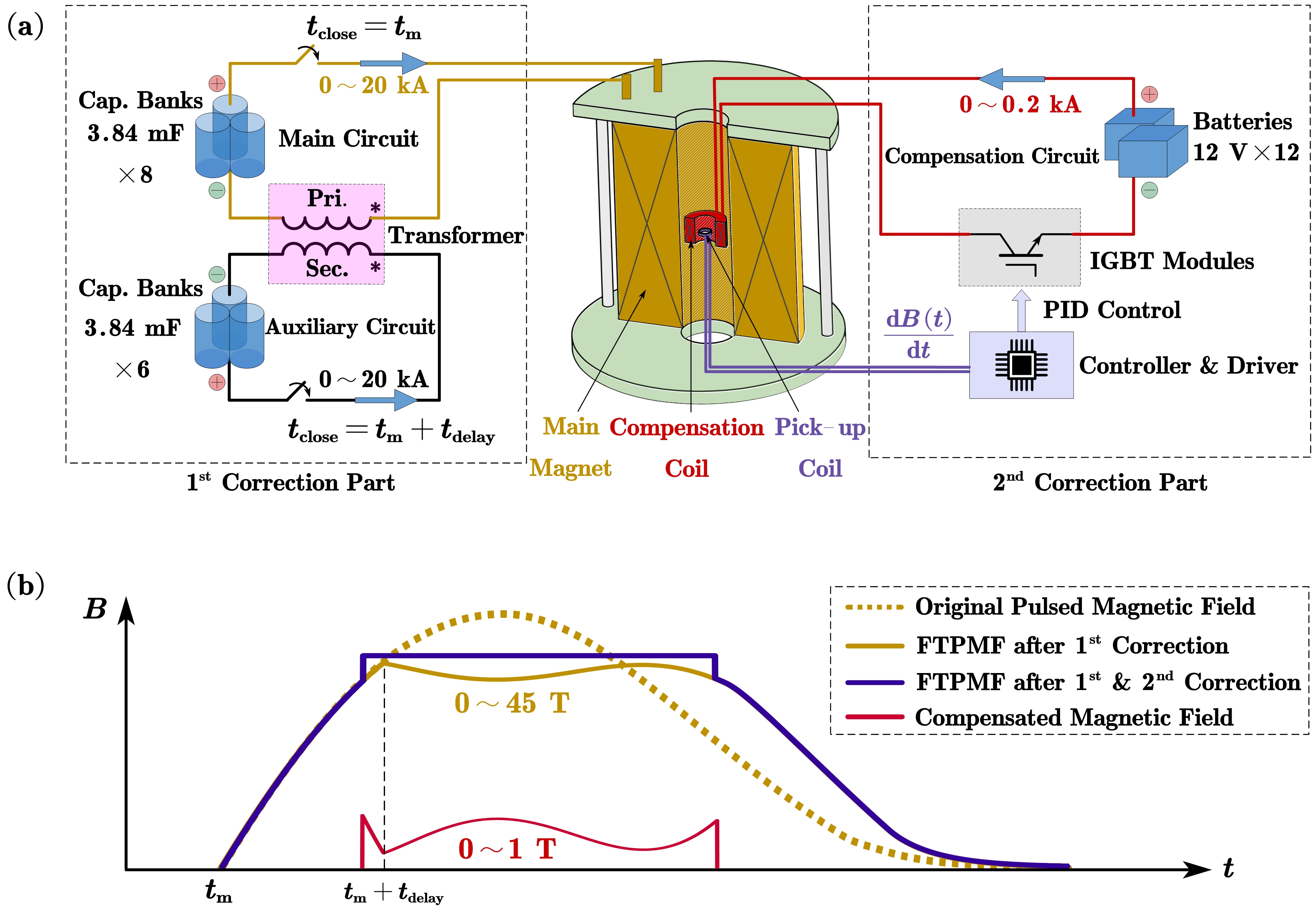}
\caption{Scheme of the two-stage corrected flat-top pulsed magnetic field for NMR measurements. (a) Diagram of the connection of main components in the two-stage corrected FTPMF system. (b) Illustration of the magnetic field waveforms generated by the two-stage corrected FTPMF system.}
\end{figure}

In terms of concrete implementation, the main magnetic field was produced by a standard 60 T pulsed magnet with a height of 150 mm and a bore of 20 mm, which has an inductance of 3.5 mH and a resistance of 30 m$\Omega$ at 77 K (the magnet was immersed in the liquid nitrogen for cooling). The main magnet was connected in series with the primary winding of a pulse transformer (refer to Ref. \cite{ref28} for specific parameters) and powered by 8 parallel capacitor banks (3.84 mF, 1.2 MJ). The secondary winding of the pulse transformer was powered by 6 parallel capacitor banks (3.84 mF, 1.2 MJ) and formed the auxiliary circuit. Before the current of the main circuit reached its peak after discharge, the auxiliary circuit was triggered to achieve the first correction. The discharge voltages of both circuits and the trigger time of the auxiliary circuit were set manually according to the optimized results of circuit simulation and experiment without the second correction. Following the different optimized parameters, the flat-top pulsed magnetic field up to 45 T after the first correction was produced, whose stability was within 0.5 T, and duration was within 12$\sim$15 ms depending on the field strength.

\begin{figure}[t]
\centering
\includegraphics[width=3.3in]{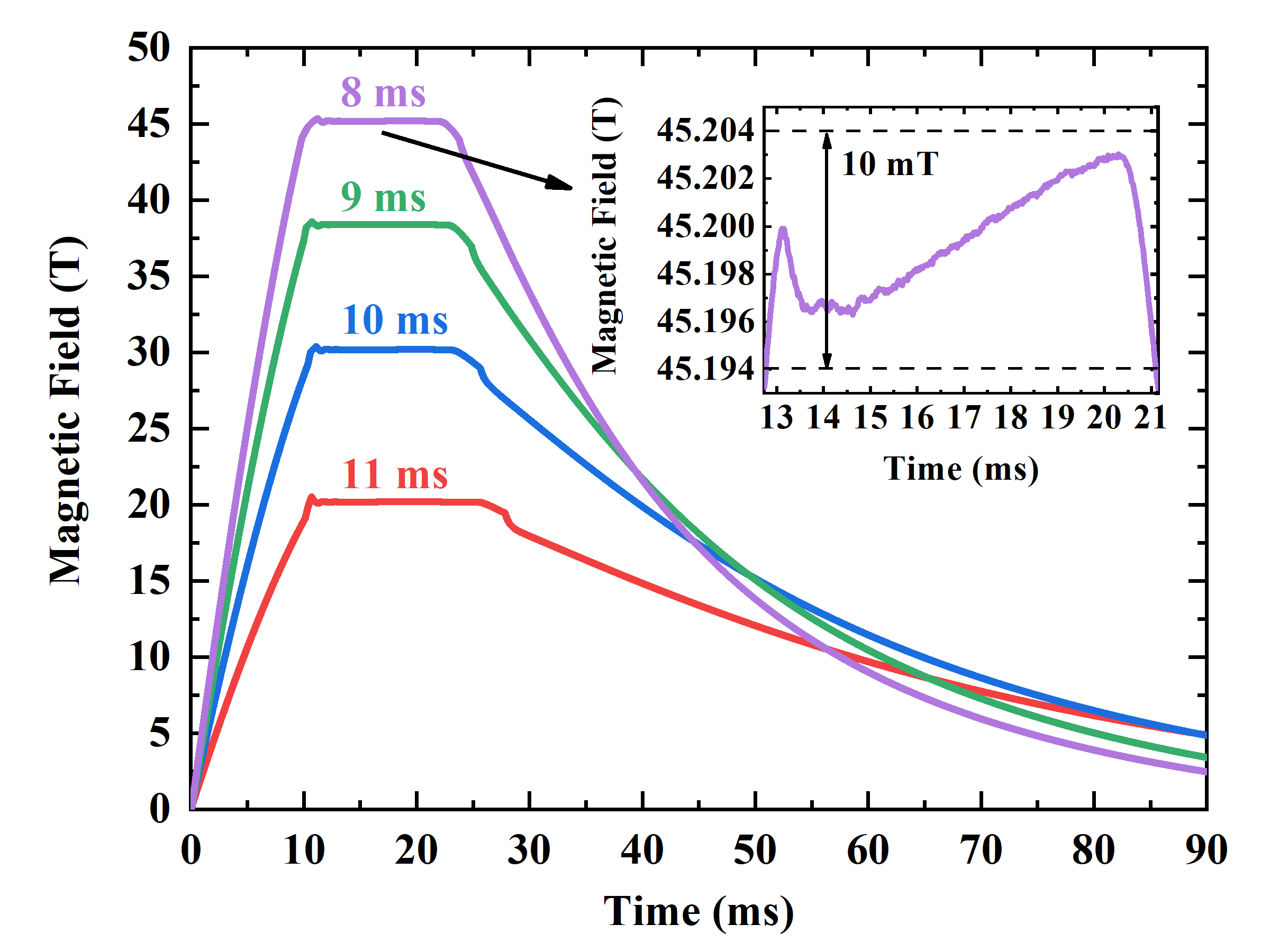}
\caption{Magnetic field profiles up to 45 T generated by the two-stage corrected flat-top system and measured by the pick-up coil. The insert shows that the magnetic field fluctuation of the flat-top was limited to less than 10 mT (integral drift had been corrected mostly).}
\end{figure}

The compensation coil used for the second correction consists of a main regulating coil with a height of 36 mm at the center of the main magnet and two oppositely wound decoupling coils with a height of 18 mm at the ends of the main regulating coil. Due to the strict radial space limitation in the main magnet bore of 20 mm, the copper compensation coil was double wound around a 17 mm Helium cryostat tail with only 0.4 mm wire diameter and reinforced by a Zylon/epoxy composite layer of 0.5 mm thickness. The compensation coil was powered by twelve 12-V lead-acid batteries, and its current was linearly regulated by insulated gate bipolar transistor (IGBT) modules with a homemade driver to generate an adjustable magnetic field from 0 to 1 T beyond 10 ms. The change rate of the magnetic field signals at the sample area was sensed by a pick-up coil based on Faraday's law of electromagnetic induction. The pick-up voltage was acquired at a sample rate of 100 kS/s by a 16-Bit analog-to-digital converter (ADC) equipped with a field programmable gate array (FPGA) module. After digital integration by the FPGA, the magnetic field strength signals were preliminarily calibrated by the electron spin resonance experiment of DPPH with an uncertainty about 2{\%}. Digital proportional integral derivative (PID) control was applied to achieve the fine compensation of the magnetic field for the second correction. The second correction part was triggered when the magnetic field strength measured by the pick-up coil was less than 1 T of the target flat-top value. Finally, the magnetic field fluctuation after the two-stage correction was limited to less than 10 mT, which is shown in Fig. 3. Due to the establishment time of the PID control and the decrease of the regulation capability of the compensation coil caused by the Joule heating, the flat-top time after the second correction was reduced by about 4 ms compared with that of the first-order correction. It is noteworthy that both centers of the main magnet and compensation coil should be aligned accurately to provide a high magnetic field homogeneity area for NMR experiments. The axial locating processes were realized by using the pick-up coil and a lock-in amplifier with an estimated error of less than 2 mm. Considering the aging of the main magnet, we would perform NMR experiments in the FTPMF up to 40 T for safety.

\subsection{Modular GHz NMR Spectrometer}
According to the properties of the FTPMF, several requirements of the NMR spectrometer are as follows: (1) Since the pulsed field is transient and its duration is typically less than 100 ms, the spectrometer should be controlled by a sophisticated timing system and has excellent transient responses. (2) The spectrometer should support high radio frequency up to several GHz, due to the fact that the resonance frequency is proportional to the field strength. For example, 2.6 GHz is needed for $^1$H at 60 T. (3) Low noise is required because the electromagnetic environment of the pulsed field is far worse than that of the DC field, and the number of times of the signal averaging is limited in the FTPMF. (4) Due to some experimental needs of sweeping the magnetic field or frequency, the RF excitation circuit should have a large RF power capacity, the RF receiving circuit also demands a broad real-time bandwidth, and the two should have good isolation. Because the above requirements cannot be completely satisfied by the commercial NMR spectrometer, a customize modular GHz NMR spectrometer was developed, in which a series of stand-alone instruments were combined for use. The advantages of the modular spectrometer are convenient for testing and maintenance, as well as further upgrading and expansion.

A diagram of the self-built modular NMR spectrometer is presented in Fig. 4. The construction of the spectrometer was based on a National Instruments (NI) PXI system including a chassis and an embedded computer, which was controlled via a self-written LabVIEW interface. Currently, the PXI system hosted three main modules, consisting of a RF generator NI PXIe-5651, a vector signal analyzer NI PXI-5661 and a pulse programmer NI PXI-6542. The RF generator can produce RF signals with frequency ranges from 500 kHz to 3.3 GHz. Correspondingly, the vector signal analyzer can support intermediate frequency (IF) down-conversion of NMR signals up to 2.7 GHz with a 20 MHz real-time bandwidth, and the NMR signals are finally recorded by a 100 MS/s and 14-Bit oscilloscope using quadrature digital down-conversion. The 100 MHz digital pulse programmer was used to receive an external trigger when the flat-top field arrived, and then trigger each independent module in the spectrometer. All components were synchronously clocked by a 10 MHz reference clock from the RF generator.

\begin{figure}[t]
\centering
\includegraphics[width=3.3in]{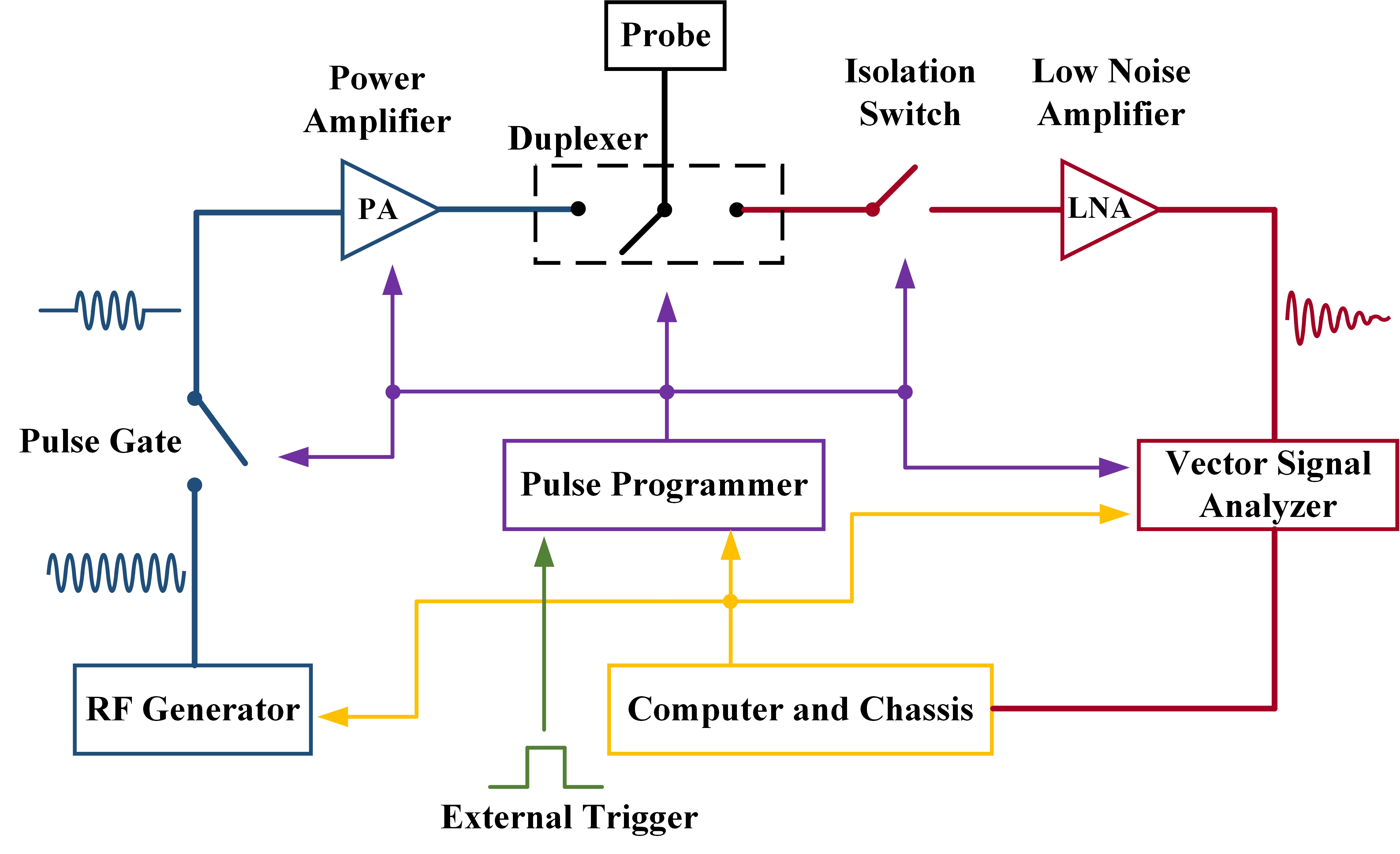}
\caption{Diagram of the modular GHz NMR spectrometer.}
\end{figure}

In the preceding stage circuit for excitation, the continuous RF signals were modulated into desired NMR sequences by a pulse gate (Analog Devices, HMC427ALP3E) with a switching time of smaller than 10 ns and isolation up to 40 dB. Then, the power of modulated signals was amplified by a Tomco 100 W RF amplifier with band 5-650 MHz. The duplexer is a key three-port component for switching between the excitation of high-power RF signals and the reception of weak NMR signals. A 1 GHz bandwidth homemade active duplexer based on PIN diodes and various impedance lines was developed with 0.56 dB insert loss, 500 W power capacity, 37 dB isolation and 1 $\upmu$s switching time. To protect the receiving part from the crosstalk of high-power excitation signals, a 70 dB isolation switch (Mini-Circuits, ZASWA-2-50DRA+) was applied with a switching time of 20 ns. A low noise amplifier N141-306CB up to 1 GHz with a gain of greater than 30 dB and noise figure of smaller than 0.8 dB was used to amplify the weak NMR signals.

\subsection{Sample Probe}
Compared with the conventional NMR probe for the DC field, the probe for the pulsed field is challenging to design because of the strict space limitation in the bore of the pulsed magnet, as well as serious electromagnetic interference and mechanical vibration environment. In practice, the bore space of the magnet available for the probe design is only 10 mm, but the probe needs to have multiple functional components, for example, including a sample chamber with enough space, a well tuned and matched resonance circuit, a temperature sensor and a heater, as well as a magnetic field pick-up coil. Because the eddy current in the probe may cause temperature rise and create additional magnetic field gradients that degrade the NMR spectral resolution, closed loops of the conducting materials perpendicular to the main magnetic field in the probe have to be avoided. In addition, all components should be well fixed.

\begin{figure}[thpb]
\centering
\includegraphics[width=3.3 in]{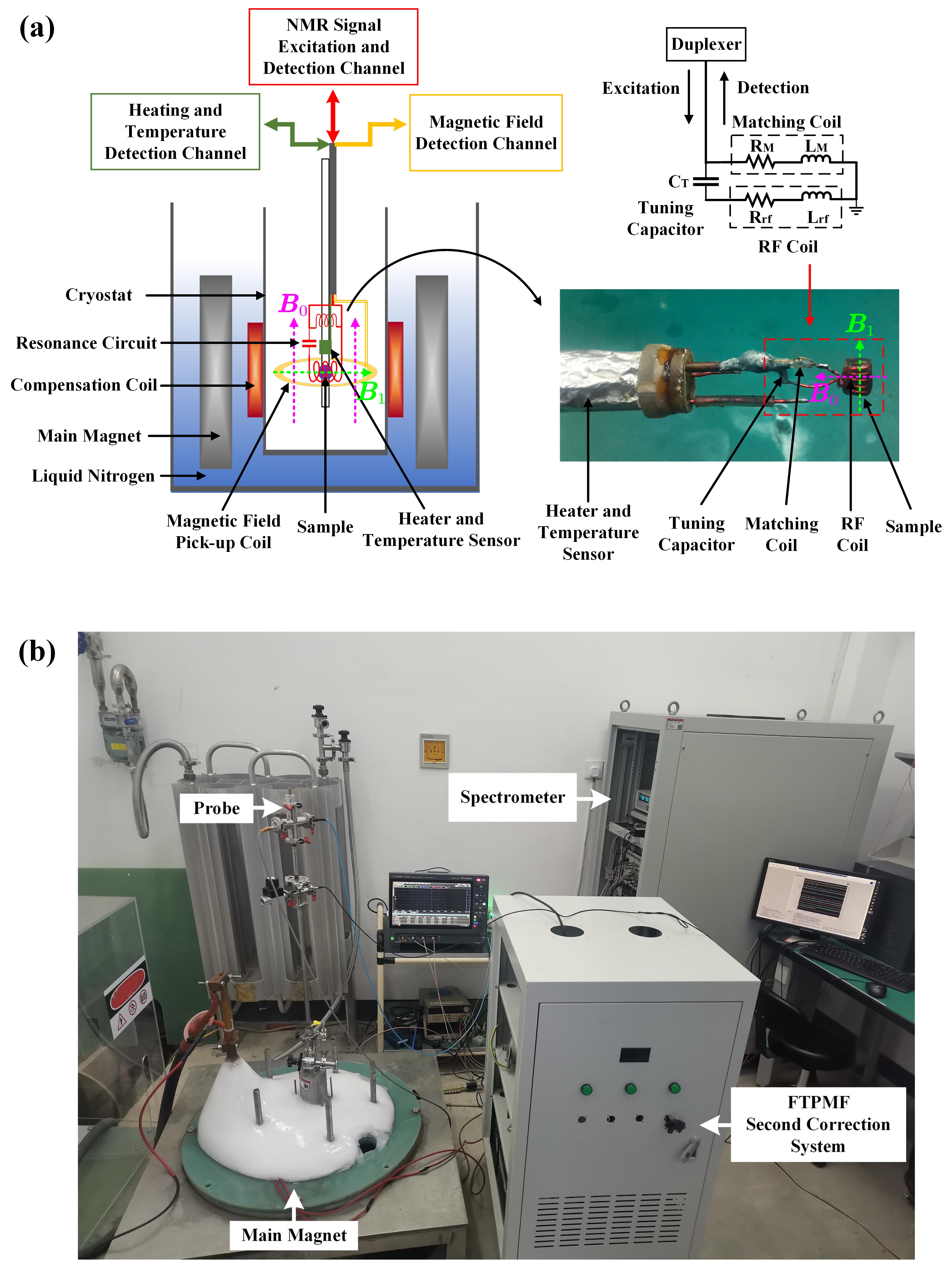}
\caption{(a) Schematic illustration of the sample probe designed for the pulsed field. The partial photo of the sample probe and connection of the resonance circuit are also shown. (b) Photo of the experimental station layout.}
\end{figure}

Fig. 5(a) shows the schematic illustration of the sample probe designed for the pulsed field. The probe was located in a cryostat that can use liquid helium or nitrogen to perform low-temperature experiments. The volume of the sample chamber was about several tens of cubic millimeters and it was surrounded by a micro RF coil to excite and detect NMR signals. A tuning capacitor and a matching coil were connected to the resonant circuit which had been preliminarily calculated and tested at room temperature. Although there was a deviation of about several MHz from the expected resonant frequency in the working condition, precise tuning was not very necessary in this pulsed field experiment. The return loss of the resonance circuit at the resonance frequency was adjusted to be less than -10 dB, meaning that more than 90\% of the power was absorbed by the RF coil. The magnetic field pick-up coil was wounded on the outer sleeve of the probe and at the same axial height as the center of the sample to ensure that the magnetic field at the sample was accurately collected. The heater and temperature sensor were located near the resonance circuit to control the local temperature around the sample. Most components were arranged vertically to improve the space utilization.

Based on the establishment of the above experimental setups, a photo of the experimental station layout is presented in Fig. 5(b). The main magnet was installed in a deep underground pit. Because the strength of the high magnetic field decays rapidly with the distance away from the magnet, the FTPMF does not heavily interfere with the surrounding electronic equipment. The experimenters are required to conduct the experiment in the other room instead of the experimental station, hence the high magnetic field does not pose a threat to the human body.

$^{93}$Nb powder (99.999\% purity) uniformly mixed by resin was selected as the testing sample. $^{93}$Nb is ideal for NMR measurements in the pulsed field, because it has a natural abundance of 100\%, a moderate gyromagnetic ratio $\gamma$=10.405 MHz/T, a short longitudinal relaxation time $T_1$ of 4.7 ms at 77 K, and a high relative sensitivity of 48{\%} compared with $^1$H. The mixed sample was cut into a cylinder and placed in the RF coil wound by the copper wire (about 2 mm in diameter, 4 mm in length, 0.3 mm wire diameter with 6 turns).  Considering the Knight shift ($K$), the resonance frequency of $^{93}$Nb is $\gamma \left( 1~+~ ^{93}K \right) B_0$, where $^{93}K= 0.87\%$.

\section{Results and Discussion}
\subsection{NMR Measurements in the DC Field and FTPMF}
 Because the signal-to-noise ratio (SNR) of a single NMR signal is generally low, it is necessary to obtain the maximum SNR as possible to ensure that the NMR signal can be clearly detected in the FTPMF. The sensitivity of the NMR signal largely depends on the power and duration of the RF pulse and needs to be well optimized\cite{ref10}. Since the pick-up coil was not well calibrated for measuring the pulsed magnetic field with an uncertainty of larger than 1{\%} and the repetition time for the pulsed magnet cooling was usually longer than 30 minutes, a pre-optimization process was preliminarily implemented in the DC field. it was carried out with our self-built spectrometer and probe in a superconducting magnet (Oxford) with high stability (better than 10 ppm/h) and homogeneity (better than 10 ppm/cm$^3$). The FID measurements of $^{93}$Nb in the DC field were performed at 14.55 T and 77 K, and the corresponding resonance frequency is 152.7 MHz. Finally, a 50 W and 1.5 $\upmu$s $\pi/2$ RF pulse was implemented to produce the FID signals. The FID signals after quadrature down-conversion are shown in Fig. 6(a), which were measured at the given RF of 152.6 MHz for the clarity of the oscillation. Considering the interference of switching noise, the starting point of the FID signals was regarded as the first zero-crossing point of the real or imaginary part. The initial sensitivity amplitude of the FID is about 8 (arbitrary unit). The decay time $T_{2}^{*}$ of the FID about 8 $\upmu$s was observed, which can be approximately regarded as the $T_2$ due to the high stability and homogeneity of the DC field. It was tested that the bandwidth of the NMR signals is about 600 kHz, corresponding to the magnetic field strength range of about 60 mT.
 
For quickly locating the NMR signals and calibrating the pick-up coil in the FTPMF, a field-sweep mode with a slope-top pulse was performed. According to the bandwidth of excitation and longitudinal relaxation time $T_1$ of $^{93}$Nb, the sweep slope rate was set to 10 mT/ms and the interval of excitation was 1 ms. After several sweep tests, the calibration process was completed and the NMR measurements could be normally carried out in the FTPMF. The FID signals at the field of 23.24 T and excited RF of 244 MHz after quadrature down-conversion are shown in Fig. 6(b). The duration of the RF pulse was reoptimized but the SNR was not significantly changed, hence we still adopted the 50 W and 1.5 $\upmu$s RF pulse. It is obvious that the decay time of the FID signals reduces from 8 $\upmu$s to 3 $\upmu$s in the FTPMF compared to the DC field, which was mainly caused by the inhomogeneity of the FTPMF and will be discussed later. Although the field strength of the FTPMF is 1.6 times higher than that of the DC field, the sensitivity of the FID does not increase due to the short polarization time of nuclei, inhomogeneity of the magnetic field and so on. These $^{93}$Nb FID results can be further confirmed by a spin-echo (SE) technique. A standard SE measurement ($\pi /2-\tau -\pi$, 1.5-20-3 $\upmu$s in this case) was performed in the FTPMF, the result of which is shown in Fig. 7. It is clear that the SE signals refocus after waiting for 20 $\upmu$s, as expected.

\begin{figure}[t]
\centering
\includegraphics[width=3.3in]{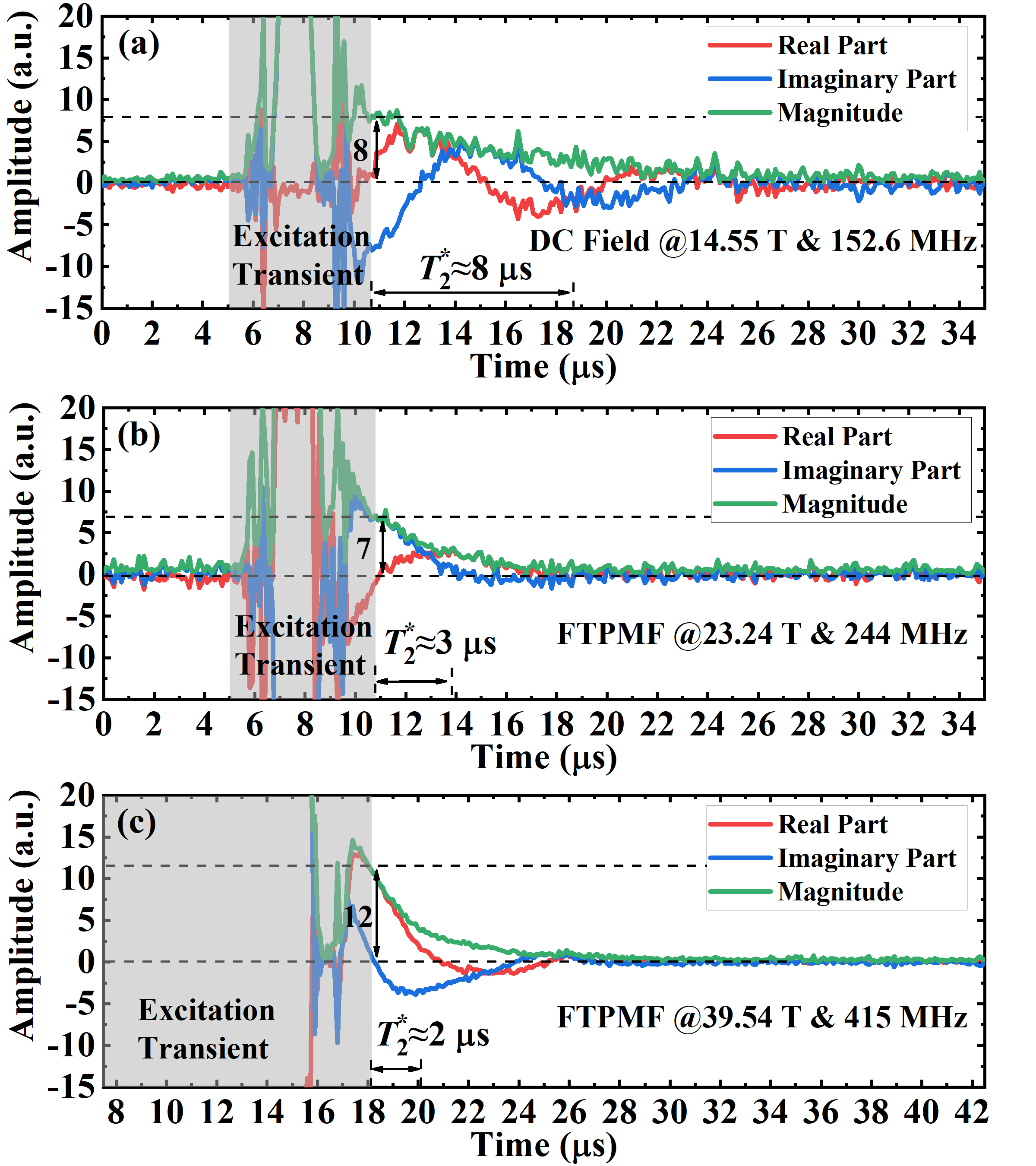}
\caption{FID signals after quadrature down-conversion: (a) at the DC field of 14.55 T and excited RF of 152.6 MHz; (b) at the FTPMF of 23.24 T and excited RF of 244 MHz. (c) at the FTPMF of 39.54 T and excited RF of 415 MHz. All measurements were carried out at a temperature of 77 K.}
\end{figure}

\begin{figure}[t]
\centering
\includegraphics[width=3.3in]{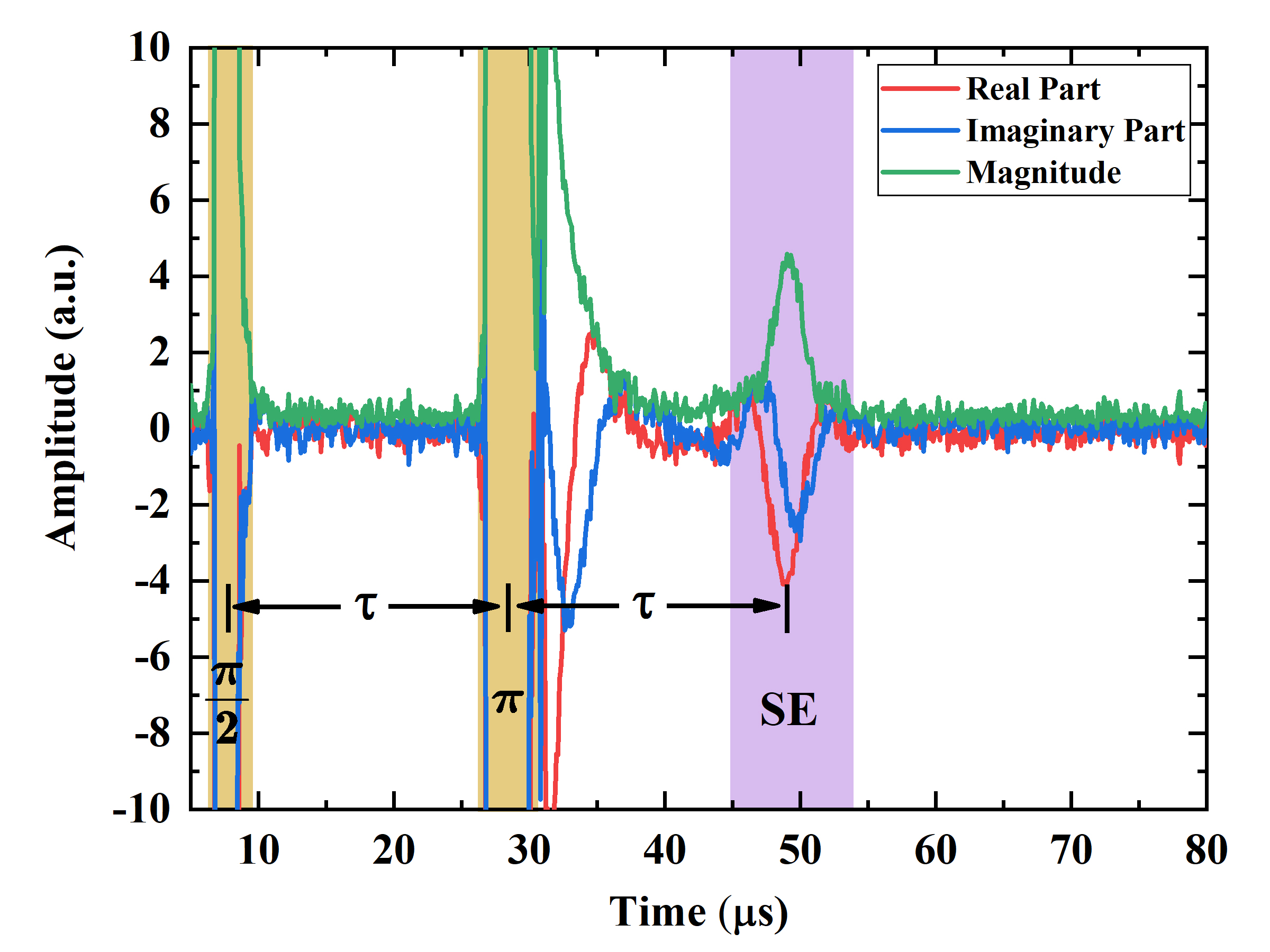}
\caption{
NMR RF pulse sequence followed by the spin-echo signals at the FTPMF of 23.26 T and excited RF of 244 MHz. The given SE sequence is 1.5-20-3.0 $\upmu$s. 
}
\end{figure}

The FID signals in the FTPMF of 39.54 T and excited RF of 415 MHz were measured and are presented in Fig. 6(c). It is noteworthy that the given $\pi/2$ RF pulse was reoptimized with the duration of 9 $\upmu$s to achieve the highest possible signal sensitivity while keeping the 50 W power unchanged. The results show that the signal sensitivity in the high field of 39.54 T increases about 1.7 times in proportion to the magnetic field strength, compared with that in the low field of 23.24 T (This can be also verified by the spectral area in Fig. 9). However, the decay time of the FID further decreases to 2 $\upmu$s due to the degradation of the magnetic field homogeneity in the absolute value.

\subsection{Stability of the FTPMF}
The NMR technique itself is also an accurate method for measuring the magnetic field, hence the FID signals were excited several times to check the stability of the FTPMF during the period of the flat-top. Fig. 8(a) shows the nine excitation points with an interval of 1 ms during the FTPMF at 39.55 T measured by the pick-up coil. The corresponding NMR spectra at the excited RF of 415 MHz are presented in Fig. 8(b). Due to the low sensitivity of the pick-up coil in the flat-top stage and the problem of the voltage drift in the ADC, the frequency offset of the NMR spectra in Fig. 8(b) does not accurately match the magnetic field strength measured by the pick-up coil in Fig. 8(a). However, The central frequency offset of the nine spectra was limited within 100 kHz, meaning the field fluctuation was less than 10 mT. A higher resolution ADC for the acquisition of the magnetic field and a smaller feedback control cycle for the second correction system are desired to further improve the stability of the FTPMF. In order to prolong the duration of the FTPMF, the main magnet and the transformer with higher inductance are applicable. However, this has to increase their volume, cost and energy consumption. Overall, the current 10 mT stability of the FTPMF in the timescale of about 10 ms would potentially find some applications for NMR experiments with short $T_1$ in condensed matter physics\cite{ref33}.

\begin{figure}[htbp]
\centering
\includegraphics[width=3.3in]{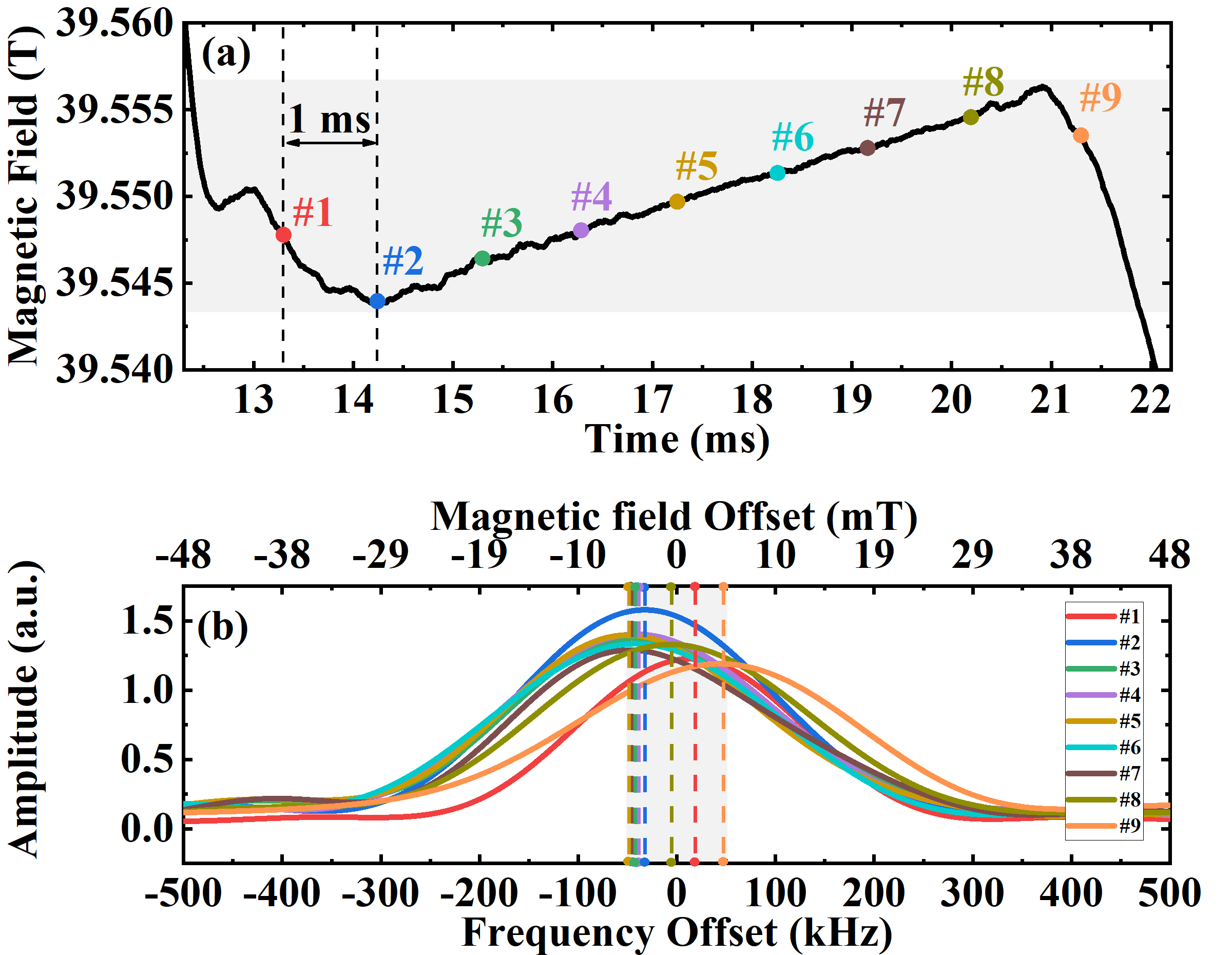}
\caption{Stability of the FTPMF checked by the NMR spectra. (a) Nine excitation points of the FID signals with an interval of 1 ms during the FTPMF at 39.55 T measured by the pick-up coil. (b) The corresponding NMR spectra at the excited RF of 415 MHz. The central frequency offset of the nine spectra was limited within 100 kHz, which corresponds well to the magnetic field fluctuation of 10 mT measured by the pick-up coil.}
\end{figure}

\subsection{Homogeneity of the FTPMF}
In section II, we revealed that the decay time $T_{2}^{*}$ of the FID signals decreases with the increase of the magnetic field spatial inhomogeneity. The inhomogeneity can lead to the broadening of the NMR spectrum, which is harmful to high-resolution NMR measurements. Compared with the commercial superconducting magnet with sophisticated shimming techniques, the pulsed magnet has great magnetic field inhomogeneity up to 1000 ppm or higher in 1 cm$^3$ space of the magnet center. Ignoring the winding error and magnet deformation, the axial distribution of the magnetic field generated by the solenoid-type magnet is approximately a quadratic function of $B_0\left( z \right) =B_0-az^2$ , where the coefficient $a>0$. Hence, small volume and highly accurate axial positioning of the sample in the pulsed magnet are important to weaken the influence of the magnetic field inhomogeneity.

The NMR spectra measured in the DC field of 14.55 T and measured in the FTPMF of 23.25 T are compared in Fig. 9. The axial position of the sample in the FTPMF was well optimized with an accuracy of less than 1 mm (labelled at 0 mm). Because of the high homogeneity of better than 10 ppm/cm$^3$ in the DC field, the FWHM of 300 ppm can be approximately attributed to the intrinsic homogeneous broadening. Although the strength of the FTPMF is higher than that of the DC field, the FWHM increases to 550 ppm due to the magnetic field inhomogeneity. Because the magnetic field inhomogeneity was caused by both the compensation coil and main magnet, the original pulsed magnetic field at the peak with a change rate of less than 1 mT/10 $\upmu$s was applied to check the inhomogeneity induced by the main magnet. The results show that the main magnet is the main source of the magnetic field inhomogeneity while the compensation coil has little impact. The spectrum of the sample positioned at 3 mm away from the optimized axial position in the pulsed field of 23.25 T was also measured and is shown in Fig. 9. It is obvious that the FWHM increases to 1440 ppm due to the heavier magnetic field inhomogeneity across the sample.

The NMR spectrum in the FTPMF of 39.55 T is also presented in Fig. 9. Despite the greater absolute broadening of the spectrum in 39.55 T compared with that in the low field of 23.25 T, the relative broadening almost remains unchanged due to the increase of resonance frequency from 244 MHz up to 415 MHz. On balance, the inhomogeneity effect of the pulsed magnet can be reduced to the level of 10$^2$ ppm with a sample volume of 10 mm$^3$ by the location optimization of the sample. A specially designed main magnet with higher magnetic field homogeneity is needed to further promote the resolution of NMR measurements\cite{ref34}.

\begin{figure}[!t]
\centering
\includegraphics[width=3.3in]{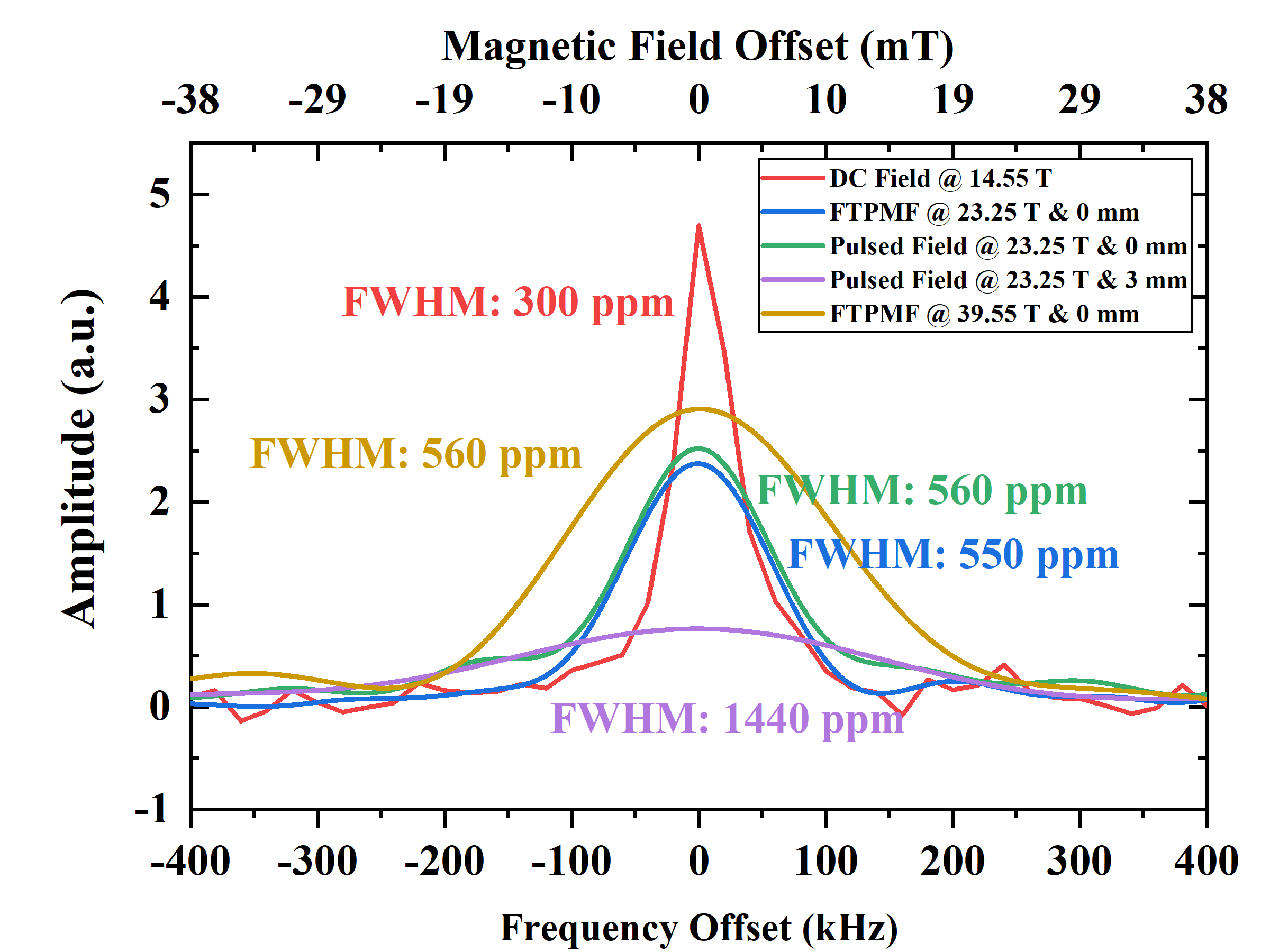}
\caption{The NMR spectra of $^{93}$Nb under different conditions. Relative FWHM is marked for comparison.}
\end{figure}

\section{Conclusion}
In this paper, we report the nuclear magnetic resonance measurements in the high flat-top pulsed magnetic field up to 40 T. The scheme of FTPMF with two-stage correction was proposed, whose magnetic field fluctuation was limited within 10 mT and duration was beyond 9 ms. Besides, the NMR spectrometer and probe suitable for the pulsed field condition were also developed. The NMR measurements of $^{93}$Nb were preoptimized in the DC field and finally carried out in the FTMPF. The results show that the stability and homogeneity of the FTPMF can reach an order of  10$^2$ ppm / 10 ms and 10$^2$ ppm / 10 mm$^3$ respectively, which is sufficient for exploring microscopic structures of some condensed matters with short $T_1$ in the high magnetic field. Crucial avenues of research in the future are the improvements of the FTPMF in many aspects including field strength, duration, stability, homogeneity and so on.

\bibliographystyle{IEEEtran}
\normalem
\bibliography{main}

\begin{thebibliography}{10}
\providecommand{\url}[1]{#1}
\csname url@samestyle\endcsname
\providecommand{\newblock}{\relax}
\providecommand{\bibinfo}[2]{#2}
\providecommand{\BIBentrySTDinterwordspacing}{\spaceskip=0pt\relax}
\providecommand{\BIBentryALTinterwordstretchfactor}{4}
\providecommand{\BIBentryALTinterwordspacing}{\spaceskip=\fontdimen2\font plus
\BIBentryALTinterwordstretchfactor\fontdimen3\font minus
  \fontdimen4\font\relax}
\providecommand{\BIBforeignlanguage}[2]{{%
\expandafter\ifx\csname l@#1\endcsname\relax
\typeout{** WARNING: IEEEtran.bst: No hyphenation pattern has been}%
\typeout{** loaded for the language `#1'. Using the pattern for}%
\typeout{** the default language instead.}%
\else
\language=\csname l@#1\endcsname
\fi
#2}}
\providecommand{\BIBdecl}{\relax}
\BIBdecl

\bibitem{ref1}
Z.~Gan, P.~Gor'kov, T.~A. Cross, A.~Samoson, and D.~Massiot, ``Seeking higher
  resolution and sensitivity for {NMR} of quadrupolar nuclei at ultrahigh
  magnetic fields,'' \emph{J. Am. Chem. Soc.}, vol. 124, no.~20, pp.
  5634--5635, 2002.

\bibitem{ref2}
P.~Van~Bentum, J.~Maan, J.~Van~Os, and A.~Kentgens, ``Strategies for
  solid-state {NMR} in high-field {Bitter} and hybrid magnets,'' \emph{Chem.
  Phys. Lett.}, vol. 376, no. 3-4, pp. 338--345, 2003.

\bibitem{ref3}
Z.~Gan, H.-T. Kwak, M.~Bird, T.~Cross, P.~Gor’kov, W.~Brey, and K.~Shetty,
  ``High-field {NMR} using resistive and hybrid magnets,'' \emph{J. Magn.
  Reson.}, vol. 191, no.~1, pp. 135--140, 2008.

\bibitem{ref4}
L.~Frydman, ``High magnetic field science and its application in the {United
  States}: {A} magnetic resonance perspective,'' \emph{J. Magn. Reson.}, no.
  242, pp. 256--264, 2014.

\bibitem{ref5}
Z.~Gan, I.~Hung, X.~Wang, J.~Paulino, G.~Wu, I.~M. Litvak, P.~L. Gor'kov, W.~W.
  Brey, P.~Lendi, J.~L. Schiano, M.~D. Bird, I.~R. Dixon, J.~Toth, G.~S.
  Boebinger, and T.~A. Cross, ``{NMR} spectroscopy up to 35.2 {T} using a
  series-connected hybrid magnet,'' \emph{J. Magn. Reson.}, vol. 284, pp.
  125--136, 2017.

\bibitem{ref6}
J.~Haase, D.~Eckert, H.~Siegel, H.~Eschrig, K.-H. M{\"u}ller, and F.~Steglich,
  ``Nuclear magnetic resonance in pulsed high-field magnets,'' \emph{Concepts
  Magn Reson Part B Magn Reson Eng}, vol.~19, no.~1, pp. 9--13, 2003.

\bibitem{ref7}
J.~Haase, F.~Steglich, D.~Eckert, D.~Siegel, H.~Eschrig, and K.~M{\"u}ller,
  ``High-field {NMR} in pulsed magnets,'' \emph{Solid State Nucl Magn Reson},
  vol.~23, no.~4, pp. 263--265, 2003.

\bibitem{ref8}
J.~Haase, D.~Eckert, H.~Siegel, H.~Eschrig, K.-H. M{\"u}ller, A.~Simon, and
  F.~Steglich, ``{NMR} at the frontier of pulsed high field magnets,''
  \emph{Physica B Condens. Matter}, vol. 346, pp. 514--518, 2004.

\bibitem{ref9}
J.~Haase, D.~Eckert, H.~Siegel, K.-H. M{\"u}ller, H.~Eschrig, A.~Simon, and
  F.~Steglich, ``{NMR} in pulsed high magnetic fields,'' \emph{J. Magn. Magn.
  Mater.}, vol. 272, pp. E1623--E1625, 2004.

\bibitem{ref10}
J.~Haase, M.~Kozlov, K.-H. M{\"u}ller, H.~Siegel, B.~B{\"u}chner, H.~Eschrig,
  and A.~Webb, ``{NMR} in pulsed high magnetic fields at 1.3 {GHz},'' \emph{J.
  Magn. Magn. Mater.}, vol. 290, pp. 438--441, 2005.

\bibitem{ref11}
J.~Haase, M.~Kozlov, A.~Webb, B.~B{\"u}chner, H.~Eschrig, K.-H. M{\"u}ller, and
  H.~Siegel, ``2 {GHz} {$^1$}{H} {NMR} in pulsed magnets,'' \emph{Solid State
  Nucl Magn Reson}, vol.~3, no.~27, pp. 206--208, 2005.

\bibitem{ref12}
B.~Meier, S.~Greiser, J.~Haase, T.~Herrmannsd{\"o}rfer, F.~Wolff-Fabris, and
  J.~Wosnitza, ``{NMR} signal averaging in 62 {T} pulsed fields,'' \emph{J.
  Magn. Reson.}, vol. 210, no.~1, pp. 1--6, 2011.

\bibitem{ref13}
J.~Kohlrautz, J.~Haase, E.~Green, Z.~Zhang, J.~Wosnitza,
  T.~Herrmannsd{\"o}rfer, H.~Dabkowska, B.~Gaulin, R.~Stern, and H.~K{\"u}hne,
  ``Field-stepped broadband {NMR} in pulsed magnets and application to
  {SrCu$_2$(BO$_3$)$_2$} at 54 {T},'' \emph{J. Magn. Reson.}, vol. 271, pp.
  52--59, 2016.

\bibitem{ref14}
E.~Abou-Hamad, P.~Bontemps, and G.~L. Rikken, ``{NMR} in pulsed magnetic
  field,'' \emph{Solid State Nucl Magn Reson}, vol.~40, no.~2, pp. 42--44,
  2011.

\bibitem{ref15}
H.~Stork, P.~Bontemps, and G.~Rikken, ``{NMR} in pulsed high-field magnets and
  application to high-{$T_{\rm C}$} superconductors,'' \emph{J. Magn. Reson.},
  vol. 234, pp. 30--34, 2013.

\bibitem{ref16}
G.-Q. Zheng, K.~Katayama, M.~Kandatsu, N.~Nishihagi, S.~Kimura, M.~Hagiwara,
  and K.~Kindo, ``{$^{59}$}{Co} {NMR} at pulsed high magnetic fields,''
  \emph{J. Low Temp. Phys.}, vol. 159, no.~1, pp. 280--283, 2010.

\bibitem{ref17}
Y.~Ihara, K.~Hayashi, T.~Kanda, K.~Matsui, K.~Kindo, and Y.~Kohama, ``Nuclear
  magnetic resonance measurements in dynamically controlled field pulse,''
  \emph{Rev. Sci. Instrum.}, vol.~92, no.~11, p. 114709, 2021.

\bibitem{ref18}
Q.~Liu, S.~Liu, Y.~Luo, and X.~Han, ``Pulsed-field nuclear magnetic resonance:
  Status and prospects,'' \emph{Matter Radiat. at Extremes}, vol.~6, no.~2, p.
  024201, 2021.

\bibitem{ref19}
J.~Toth and S.~Bole, ``Design, construction, and first testing of a 41.5 {T}
  all-resistive magnet at the {NHMFL} in {Tallahassee},'' \emph{IEEE Trans.
  Appl. Supercond.}, vol.~28, no.~3, pp. 1--4, 2017.

\bibitem{ref20}
J.~Liu, Q.~Wang, L.~Qin, B.~Zhou, K.~Wang, Y.~Wang, L.~Wang, Z.~Zhang, Y.~Dai,
  H.~Liu, X.~Hu, H.~Wang, C.~Cui, D.~Wang, H.~Wang, J.~Sun, W.~Sun, and
  L.~Xiong, ``World record 32.35 tesla direct-current magnetic field generated
  with an all-superconducting magnet,'' \emph{Supercond Sci Technol}, vol.~33,
  no.~3, p. 03LT01, 2020.

\bibitem{ref21}
S.~Hahn, K.~Kim, K.~Kim, X.~Hu, T.~Painter, I.~Dixon, S.~Kim, K.~R. Bhattarai,
  S.~Noguchi, J.~Jaroszynski, and D.~C. Larbalestier, ``45.5-tesla
  direct-current magnetic field generated with a high-temperature
  superconducting magnet,'' \emph{Nature}, vol. 570, no. 7762, pp. 496--499,
  2019.

\bibitem{ref22}
R.~Battesti, J.~Beard, S.~Böser, N.~Bruyant, D.~Budker, S.~A. Crooker, E.~J.
  Daw, V.~V. Flambaum, T.~Inada, I.~G. Irastorza, F.~Karbstein, D.~L. Kim,
  M.~G. Kozlov, Z.~Melhem, A.~Phipps, P.~Pugnat, G.~Rikken, C.~Rizzo,
  M.~Schott, Y.~K. Semertzidis, H.~H. ten Kate, and G.~Zavattini, ``High
  magnetic fields for fundamental physics,'' \emph{Phys. Rep.}, vol. 765, pp.
  1--39, 2018.

\bibitem{ref23}
M.~Jaime, R.~Daou, S.~A. Crooker, F.~Weickert, A.~Uchida, A.~E. Feiguin, C.~D.
  Batista, H.~A. Dabkowska, and B.~D. Gaulin, ``Magnetostriction and magnetic
  texture to 100.75 {Tesla} in frustrated {SrCu$_2$(BO$_3$)$_2$},'' \emph{Proc.
  Natl. Acad. Sci. U.S.A.}, vol. 109, no.~31, pp. 12\,404--12\,407, 2012.

\bibitem{ref24}
Y.~Kohama, T.~Nomura, S.~Zherlitsyn, and Y.~Ihara, ``Time-resolved measurements
  in pulsed magnetic fields,'' \emph{J. Appl. Phys.}, vol. 132, no.~7, p.
  070903, 2022.

\bibitem{ref25}
B.~Meier, J.~Kohlrautz, J.~Haase, M.~Braun, F.~Wolff-Fabris, E.~Kampert,
  T.~Herrmannsd{\"o}rfer, and J.~Wosnitza, ``Nuclear magnetic resonance
  apparatus for pulsed high magnetic fields,'' \emph{Rev. Sci. Instrum.},
  vol.~83, no.~8, p. 083113, 2012.

\bibitem{ref26}
F.~Weickert, B.~Meier, S.~Zherlitsyn, T.~Herrmannsd{\"o}rfer, R.~Daou,
  M.~Nicklas, J.~Haase, F.~Steglich, and J.~Wosnitza, ``Implementation of
  specific-heat and {NMR} experiments in the 1500 ms long-pulse magnet at the
  {Hochfeld-Magnetlabor Dresden},'' \emph{Meas Sci Technol}, vol.~23, no.~10,
  p. 105001, 2012.

\bibitem{ref27}
F.~Jiang, T.~Peng, H.~Xiao, J.~Zhao, Y.~Pan, F.~Herlach, and L.~Li, ``Design
  and test of a flat-top magnetic field system driven by capacitor banks,''
  \emph{Rev. Sci. Instrum.}, vol.~85, no.~4, p. 045106, 2014.

\bibitem{ref28}
S.~Wang, T.~Peng, F.~Jiang, S.~Jiang, S.~Chen, L.~Deng, R.~Huang, and L.~Li,
  ``Upgrade of the pulsed magnetic field system with flat-top at the {WHMFC},''
  \emph{IEEE Trans. Appl. Supercond.}, vol.~30, no.~4, pp. 1--4, 2020.

\bibitem{ref29}
Y.~Kohama and K.~Kindo, ``Generation of flat-top pulsed magnetic fields with
  feedback control approach,'' \emph{Rev. Sci. Instrum.}, vol.~86, no.~10, p.
  104701, 2015.

\bibitem{ref30}
L.~Campbell, H.~Boenig, D.~Rickel, J.~Schillig, H.~Schneider-Muntau, and
  J.~Sims, ``The {NHMFL} long-pulse magnet system- 60--100 {T},'' \emph{Physica
  B Condens. Matter}, vol. 216, no. 3-4, pp. 218--220, 1996.

\bibitem{ref31}
S.~Zhang, Z.~Wang, T.~Ding, H.~Xiao, J.~Xie, and X.~Han, ``Realization of
  high-stability flat-top pulsed magnetic fields by a bypass circuit of {IGBTs}
  in the active region,'' \emph{IEEE Trans. Power Electron.}, vol.~35, no.~3,
  pp. 2436--2444, 2019.

\bibitem{ref32}
D.~Li, H.~Ding, Y.~Fang, S.~Zhang, and D.~Pan, ``Generation of a flat-top
  magnetic field with multiple-capacitor power supply,'' \emph{IEEE Access},
  vol.~10, pp. 35\,550--35\,560, 2022.

\bibitem{levitt2013spin}
M.~H. Levitt, \emph{Spin dynamics: basics of nuclear magnetic resonance}.\hskip
  1em plus 0.5em minus 0.4em\relax John Wiley \& Sons, 2013.

\bibitem{slichter2013principles}
C.~P. Slichter, \emph{Principles of magnetic resonance}.\hskip 1em plus 0.5em
  minus 0.4em\relax Springer Science \& Business Media, 2013, vol.~1.

\bibitem{chavhan2009principles}
G.~B. Chavhan, P.~S. Babyn, B.~Thomas, M.~M. Shroff, and E.~M. Haacke,
  ``Principles, techniques, and applications of {T2*}-based {MR} imaging and
  its special applications,'' \emph{Radiographics}, vol.~29, no.~5, pp.
  1433--1449, 2009.

\bibitem{ref33}
M.~D. Bird, W.~W. Brey, T.~A. Cross, I.~R. Dixon, A.~Griffin, S.~T. Hannahs,
  J.~Kynoch, I.~M. Litvak, J.~L. Schiano, and J.~Toth, ``Commissioning of the
  36 {T} series-connected hybrid magnet at the {NHMFL},'' \emph{IEEE Trans.
  Appl. Supercond.}, vol.~28, no.~3, pp. 1--6, 2017.

\bibitem{ref34}
A.~Orlova, P.~Frings, M.~Suleiman, and G.~Rikken, ``New high homogeneity 55 {T}
  pulsed magnet for high field {NMR},'' \emph{J. Magn. Reson.}, vol. 268, pp.
  82--87, 2016.

\end{thebibliography}

\end{document}